\documentclass[twocolumn]{aastex7}
\usepackage{comment}
\usepackage{xcolor}
\usepackage{commath}
\usepackage{soul}
\usepackage{CJKutf8}
\usepackage[normalem]{ulem}

\def\mso{M_\odot}
\def\kms{\mathrm{km\,s}^{-1}}

\begin{document}

\begin{CJK*}{UTF8}{gbsn}

\title{Can isolated binaries form unequal-mass binary black-hole mergers with a high-spin primary black hole?}

\author[orcid=0000-0001-9565-9462,sname='Xu']{Xiao-Tian Xu (徐啸天)}
\affiliation{Tsung-Dao Lee Institute, Shanghai Jiao-Tong University, 1 Lisuo Road, Shanghai 201210, People’s Republic of China}
\email[show]{xxu-tdli@sjtu.edu.cn}

\author[orcid=0000-0001-9565-9462,sname='Lai']{Dong Lai}
\affiliation{Tsung-Dao Lee Institute, Shanghai Jiao-Tong University, 1 Lisuo Road, Shanghai 201210, People’s Republic of China}
\affiliation{Center for Astrophysics and Planetary Science, Department of Astronomy, Cornell University, Ithaca, NY 14853, USA}
\email[show]{donglai@sjtu.edu.cn}  

\author[orcid=0000-0002-0643-8295,sname='Liu']{Bin Liu}
\affiliation{Institute for Astronomy, School of Physics, Zhejiang University, 310058 Hangzhou, People’s Republic of China}
\affiliation{Center for Cosmology and Computational Astrophysics, Institute for Advanced Study in Physics, Zhejiang University, 310027 Hangzhou, People’s Republic of China}
\email[show]{liubin23@zju.edu.cn}

\begin{abstract}
GWTC-5 has revealed a subpopulation of merging binary black holes (BBHs) with a high-spin primary black hole (BH) and possibly unequal BH masses. GW241110 additionally exhibits a large spin-orbit misalignment, which suggests a hierarchical-merger origin. However, other formation scenarios are possible or even likely, especially for events without constraints on spin-orbit misalignment. As an alternative to hierarchical mergers, we investigate whether binary evolution can produce unequal-mass BBH mergers with a high-spin primary BH. Rather than performing comprehensive population-synthesis calculations, we examine the evolutionary pathways of forming merging BBHs and assess their uncertainties. We identify two possible pathways for producing unequal-mass BBHs with a high-spin primary. In initially wide binaries, mass-ratio reversal can make the tidally spun-up second-born BH both more massive and more rapidly rotating than the first-born BH; alternatively, in an initially close, unequal-mass binary, the primary star may evolve chemically homogeneously, while the secondary star evolves normally, producing a high-spin first-born BH that is massive than its companion.  Generally, large spin-orbit misalignment can be produced by large natal kicks or tertiary-induced nodal precession and/or Zeipel-Lidov-Kozai oscillations.  We conclude that hierarchical mergers are not uniquely required to produce unequal-mass BBHs with a high-spin primary BH, although each isolated-binary pathway faces important theoretical constraints. Future detections of more merger events with primary BH spins around 0.7 would discriminate between binary evolution and hierarchical merger origin.
\end{abstract}

\keywords{ 
\uat{Binary stars}{154}  --- 
\uat{Stellar astronomy}{1583} --- \uat{Gravitational wave sources}{677} --- \uat{Massive stars}{732}
}

\received{}
\revised{}
\accepted{}

\submitjournal{ApJ}

\section{Introduction}

Since the detection of the first gravitational-wave (GW) event \citep{Abbott2016PhRvL.116f1102A}, the LIGO-Virgo-KAGRA (LVK) Collaboration has detected GW events at an accelerated rate. In the recently released fifth  Gravitational-Wave Transient Catalogue \citep[GWTC-5;][]{LVK2026arXiv260527223T,LVK2026arXiv260527224T,LVK2026arXiv260527225T}, nearly 400 events are reported, most of which originate from the mergers of two stellar-mass black holes (BHs), and some features of this merging binary black hole (BBH) population have started to emerge. In particular, the LVK Collaboration reports a population characterised by a high-spin primary BH  \citep{LVK2026arXiv260527226T}. While the overall mass-ratio distribution of this high-spin population is not well constrained, some of the individual events show strong evidence for unequal masses, with mass ratios ($q$) around 0.5 \citep{Abac2025ApJ...993L..21A,LVK2026arXiv260527225T}. Among these, GW241011 and GW241110 stand out for their precisely measured primary spins ($a_1$) \citep[GW241011: $a_1=0.78^{+0.09}_{-0.09}$ and $q=0.30^{+0.09}_{-0.08}$; GW241110: $a_1=0.61^{+0.33}_{-0.40}$ and $q=0.45^{+0.32}_{-0.17}$;][]{Abac2025ApJ...993L..21A}, and GW241110 also exhibits a large spin-orbit misalignment \citep{Abac2025ApJ...993L..21A}. 

Stellar-born BHs are usually expected to have negligible spins due to the efficient core-envelope angular momentum transport during stellar evolution \citep{Fuller2019ApJ...881L...1F}. Differential rotation inside stars can generate magnetic fields through the Tayler--Spruit dynamo, which efficiently transfers angular momentum from the helium core of an evolved star to the rapidly expanding hydrogen-rich envelope. Consequently, the helium core only retains a small amount of angular momentum at core collapse, leading to a negligible natal BH spin\footnote{It has been proposed that weak core-envelope coupling can reproduce the high spins in BH high-mass X-ray binaries \citep{Qin2019}. However, it is unclear whether these BHs are truly fast rotating \citep{Zdziarski2026NewAR.10201746Z}.}. On the other hand, an equal-mass BBH merger with negligible initial spins results in a second-generation BH having a spin parameter around 0.7 \citep{Hinder2018PhRvD..98d4015H}. 
In a dense stellar environment, such a second-generation BH may dynamically pair with another BH \citep{Antonini2016ApJ...831..187A,Rodriguez2019PhRvD.100d3027R,Gerosa2021NatAs...5..749G,Liu2026ApJ..1008...66L}, potentially producing an unequal-mass BBH. However, the efficiency of a second-generation BH pairing with a much less first-generation BH remains uncertain and depends on the host environment \citep[e.g.,][]{Liu2026ApJ..1008...66L}. The combination of the low mass ratio and the high primary BH spin therefore motivates the hierarchical-merger interpretations for both GW241011 and GW241110 \citep{Abac2025ApJ...993L..21A}. Dynamical assembly also produces a wide range of spin-orbit misalignment, making the large misalignment inferred for GW241110 additional support for the hierarchical-merger scenario \citep{Abac2025ApJ...993L..21A}.
However, other scenarios are not excluded, especially for events without constraints on the spin-orbit misalignment \citep[e.g.,][]{Llobera-Querol2026arXiv260405492L}.

As an alternative to the hierarchical-merger scenario, we investigate possible formation channels in isolated binaries for unequal-mass BBH mergers with a high-spin primary BH, motivated by the inferred high-spin BH population in GWTC-5. Rather than performing a comprehensive population-synthesis study, in Sect.\,\ref{sec:channels} we examine several plausible evolutionary pathways and discuss their uncertainties and limitations, and thus provide a useful framework for interpreting the high-spin BHs detected through GW observations. In Sect.\,\ref{sec:misalignment}, we explore the possibilities of producing large spin-orbit misalignments by BH kicks or by a tertiary companion. We summarise our work in Sect.\,\ref{sec:conclusion}.

\section{Forming high-spin primary black holes in isolated binaries\label{sec:channels}}

While the majority of massive stars have a nearby companion \citep{Sana2012,Sana2013A&A...550A.107S,Moe2017ApJS..230...15M,Sana2025}, only a small fraction is expected to produce merging BBHs \citep[e.g., Chapter 3 in][]{Xu2024thesis}. So far, three formation channels in isolated binaries have been identified, which are named after their key evolutionary processes: the common envelope evolution channel (CEE channel), stable mass transfer channel (SMT channel), and chemically homogeneous evolution channel (CHE channel). The CEE channel works in initially wide binaries, while the SMT channel can occur in both initially wide and close binaries. The CHE channel requires an orbit close enough to trigger strong rotational mixing through tidal spin-up and deformation.
We refer to \citet{Langer2012}, \citet{Han2020RAA....20..161H}, and \citet{Marchant2024ARA&A..62...21M} for comprehensive reviews on the evolution of massive stars and binary stars. In the following, we discuss how isolated binaries may produce unequal-mass BBH mergers with a high-spin primary BH for three initial conditions: wide binaries (Sect.\,\ref{sec:wide-binaries}), close binaries (Sect.\,\ref{sec:close-binaries}), and chemically homogeneous stars (Sect.\,\ref{sec:che}). 

To avoid ambiguity, we clarify the terms used in the following discussion. We use `initial primary' and `initial secondary' for the more and less massive stars at zero-age main sequence. `First-born' and `second-born' BH refer to the order of BH formation along binary evolution, whereas `primary' and `secondary' BH refer to the more and less massive BH in a merging BBH. We note that the first-born BH needs not to be the primary BH. Correspondingly, the observed mass ratio of a merging BBH should not exceed 1 (secondary BH over primary BH), while the mass ratio of the second-born BH over the first-born BH  can exceed 1.  

\subsection{Wide binaries\label{sec:wide-binaries}}

\begin{figure*}[t]
    \centering
    \includegraphics[trim=0.8cm 2.5cm 3.5cm 0cm,width=0.95\linewidth]{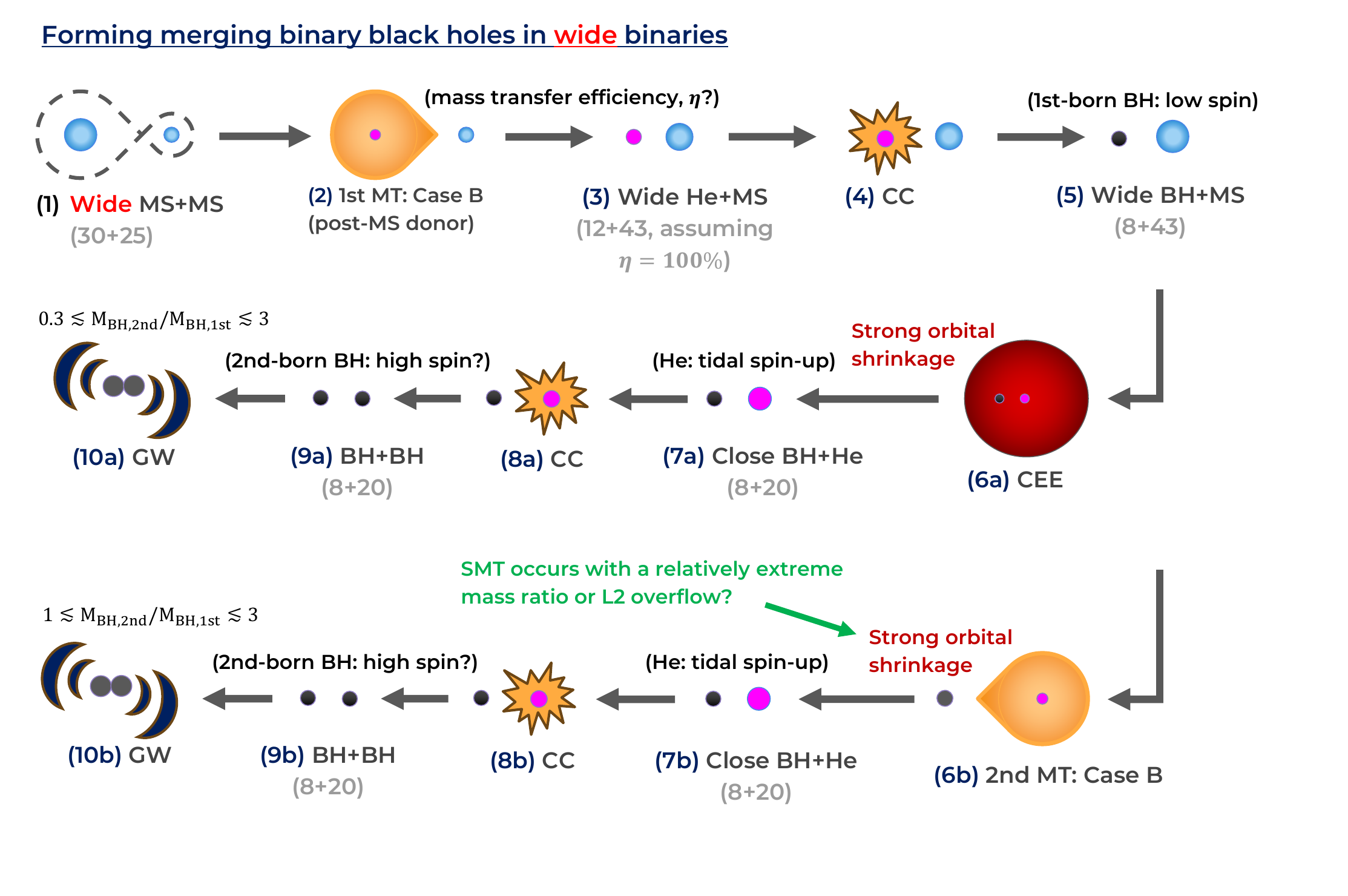}
    \caption{Formation paths of merging binary black holes in initially wide binaries. The system begins with (1) a wide binary containing two main-sequence (MS) stars. The fast expansion of the primary star after core-hydrogen depletion triggers (2) a stable mass-transfer (MT) phase. Here, `Case\,B' denotes MT from a hydrogen-shell-burning mass donor \citep{Kippenhahn1967ZA.....65..251K}. The mass donor is stripped to its helium (He) core, forming (3) a long-period He+MS binary. The He star produces a BH through (4) core collapse (CC), leading to (5) a long-period BH+MS binary. The formation of a merging BBH requires strong orbital shrinkage during the MT triggered by the expansion of the initial secondary star (see text). This can occur either through the common envelope evolution (CEE; 6a) or stable MT (6b). Both pathways produce (7a,b) a tight BH+He binary. During this stage, tides can spin up the He star. Through the second CC event (8a,b), the system may form (9a,b) a BBH with a high-spin second-born BH, which may (10a,b) merge within the Hubble time. 
     The values in grey parentheses show the component masses, in $M_\odot$, for an evolutionary example (see text). Here, $M_{\rm BH,1st}$ and $M_{\rm BH,2nd}$ denote the masses of the first- and second-born BHs, and the ranges of $M_{\rm BH,2nd}/M_{\rm BH,1st}$ are intended only as approximate references. We refer to \citet{Broekgaarden2022ApJ...938...45B} and \citet{Olejak2024A&A...689A.305O} for population synthesis predictions.
}
    \label{fig:wide}
\end{figure*}

Figure\,\ref{fig:wide} illustrates the evolutionary paths of merging BBHs for initially wide binaries, which require strong orbital shrinkage (step 6a,b in Fig.\,\ref{fig:wide}) to bring the binary components close enough for the resulting BBH to merge within the Hubble time. The mechanisms driving this orbital shrinkage are discussed below. In the following close BH+helium (He)-star stage (step 7a,b in Fig.\,\ref{fig:wide}), efficient tidal spin-up of the He star may occur, producing a high-spin second-born BH \citep[e.g.,][]{Kushnir2016MNRAS.462..844K,Qin2019,Fuller2022MNRAS.511.3951F,Bavera2023NatAs...7.1090B}. 

Since more massive stars evolve faster, the first-born BH is usually expected to be the observed primary BH. However, the birth order can be reversed in some cases (i.e., the second-born BH becomes the observed primary BH), a process often referred to as `mass-ratio reversal' \citep{Broekgaarden2022ApJ...938...45B,smith2026massqueradeimpactsmassratio,Hu2026}, allowing the production of high-spin primary BHs. This occurs if a large amount of material is ejected during the formation of the first-born BH\footnote{It remains uncertain which stars can form BH \citep[e.g.,][]{Sukhbold2016,Chieffi2020ApJ...890...43C,Burrows2025ApJ...987..164B}. Whether a star explodes may also be affected by binary interaction \citep{Schneider2021}. }\citep[i.e., step 4 in Fig.\,\ref{fig:wide};][]{Antoniadis2022A&A...657L...6A,smith2026massqueradeimpactsmassratio}, or if the initial secondary star accretes a large fraction of the transferred material during the first MT phase \citep[step 2 in Fig.\,\ref{fig:wide};][]{Broekgaarden2022ApJ...938...45B,Olejak2024A&A...689A.305O,smith2026massqueradeimpactsmassratio}. In the latter case, it is usually assumed that the convective core of the MS mass gainer grows with mass accretion. This is motivated by observational constraints favouring relatively efficient semiconvection \citep{Langer1991,Abel2019}, which allows accreted material to overcome the chemical gradient barrier above the convective core and mix into the core \citep[i.e., rejuvenation;][]{Braun1995A&A...297..483B}. This accretion-enhanced convective core leads to a larger second-born BH mass than what would be produced by the original MS secondary.

In Fig.\,\ref{fig:wide}, we also give an evolutionary example of forming an unequal-mass BBH merger with a high-spin primary BH (numbers in grey parentheses in Fig.\,\ref{fig:wide}). The masses of the He cores are estimated from the detailed models in \citet{Xu2025arXiv250323876X}.
The initial binary contains a $30\mso$ primary and a $25\mso$ secondary. The primary star develops a He core of $12\mso$, resulting in a BH of $8\,\mso$ \citep{EggenbergerAndersen2026arXiv260501405E}. Assuming a conservative mass transfer, the secondary star grows to $43\mso$. The rejuvenated star forms a He core of about $20\mso$, and its H-rich envelope is stripped during the second MT phase. If the orbital period of the BH+He stage is short enough, the second-born BH ($20\mso$) can have a spin as large as 1 \citep{Kushnir2016MNRAS.462..844K,Qin2019,Ma2023ApJ...952...53M}.  

The key process in this evolutionary path is the strong orbital shrinkage required to produce close BH+He binary (step 7a,b in Fig.\,\ref{fig:wide}) during the second MT. This orbital shrinkage naturally occurs in the CEE, where the mass gainer spirals into the envelope of the mass donor, and orbital energy unbinds the envelope material \citep{Ivanova2013,Ropke2023LRCA....9....2R}. However, the theoretical merger rates from the CEE channel are highly sensitive to how this process is treated \citep[e.g.,][]{Broekgaarden2026arXiv260605322B}, and both the conditions of initiating and surviving the CEE remain uncertain.  
Alternatively, the required orbital shrinkage may also occur if the second MT is stable (step 6b in Fig.\,\ref{fig:wide}). 
Contrary to earlier expectation \citep[e.g.,][]{Soberman1997}, recent studies of mass transfer stability suggest that SMT can take place with a convective donor or a mass ratio (mass gainer / mass donor) below $\sim0.3$ \citep{Pavlovskii2017MNRAS.465.2092P,Ge2020,Marchant2021,Gallegos-Garcia2021,Picco2024A&A...681A..31P}
Meanwhile, a lower mass ratio leads to stronger orbital shrinkage during SMT, 
enabling wide BH+OB binaries with relatively extreme mass ratios to form merging BBHs through SMT \citep{vandenHeuvel2017MNRAS.471.4256V,vanSon2022ApJ...940..184V,Marchant2021,Gallegos-Garcia2021,Olejak2024A&A...689A.305O,Klencki2026A&A...706A.296K}. In addition, strong orbital shrinkage may also occur if mass escapes from the system through the outer Lagrangian point \citep{Olejak2024A&A...689A.305O,Klencki2026A&A...706A.296K}. 

Whether the second MT is stable or unstable, the evolution of initially wide binaries expects a population of of long-period, circular binaries containing an OB-type star and a BH (step 5 in Fig.\,\ref{fig:wide}) or an OB-type star and a WR star companion\footnote{Classical Wolf--Rayet (WR) stars are expected to be massive He stars \citep{Langer2012}, while some WR stars are likely core-hydrogen-burning stars \citep{Grafener2008A&A...482..945G}} \citep[step 3 in Fig.\,\ref{fig:wide}; e.g.,][]{Langer2020,Xu2025arXiv250323876X,Schurmann2025arXiv250323878S}.  
Detecting such binaries would provide insight into the evolution of initially wide massive binaries. 
Long-period post-mass-transfer systems containing a neutron star primary, such as Be X-ray binaries, have already been detected \citep[e.g.,][]{Haberl2016}. 
However, clear candidate systems for wide BH+OB binaries are still observationally missing \citep[e.g.,][]{Janssens2023A&A...677L...9J,Bodensteiner2025A&A...698A..38B,El-Badry2025OJAp....8E.128E}. 
Long-period WR+OB binaries are also observationally rare \citep{Schootemeijer2024A&A...689A.157S,Deshmukh2024A&A...692A.109D}, particularly in low-metallicity environments \citep{Schootemeijer2024A&A...689A.157S}, where massive BHs are expected to form \citep{Belczynski2016,Marchant2016}. Because these key evolutionary stages (steps 3 and 5 in Fig.\,\ref{fig:wide}) are observationally rare or missing, the mass transfer efficiency in wide massive binaries remains uncertain (step 2 in Fig.\,\ref{fig:wide}), and even whether the primary star can fill its Roche lobe in these systems has also become questionable \citep{Olejak2025arXiv251110728O,Pauli2026A&A...707A..11P,Xu2026arXiv260314840X}.

\subsection{Close binaries: stable mass transfer\label{sec:close-binaries}}

\begin{figure*}[t]
    \centering
    \includegraphics[trim=0.7cm 2.5cm 2.7cm 0cm,clip,width=0.95\linewidth]{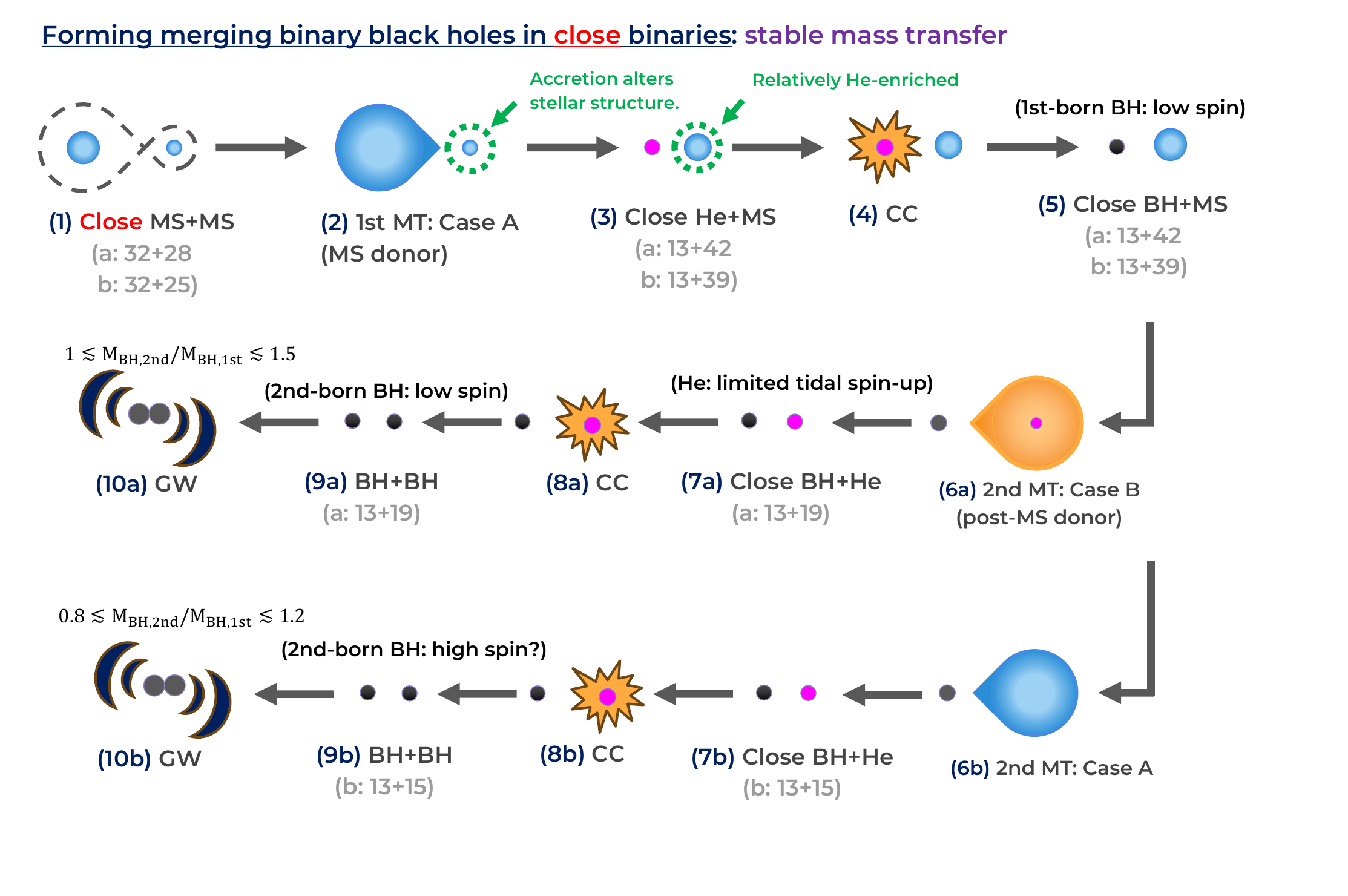}
    \caption{Formation paths of merging binary black holes in initially close binaries through stable mass transfer. The system begins with (1) a close binary containing two main-sequence (MS) stars. The primary star slowly expands during its MS evolution, (2) filling the Roche lobe. Here, `Case\,A' denotes MT from a core-hydrogen-burning mass donor, whereas `Case\,B' denotes MT from a hydrogen-shell-burning donor \citep{Kippenhahn1967ZA.....65..251K}. After the MT phase, the system becomes (3) a short-period He+MS binary, in which the MS companion is relatively helium enriched because of the prior mass accretion. Through (4) the first core-collapse (CC) event, the system produces (5) a close BH+MS binary, where the MS star is likely an OB-type star. Then it divides into (6a,b) two branches depending on the evolutionary stage of the companion star at the onset of the second MT phase (i.e., Case\,A or Case\,B). In either case, the accretion-induced structural changes during the first MT phase (2) affect the stability of the second MT phase \citep{Xu2025arXiv251220054X}. Both paths then evolve through (7a,b) a close BH+He stage --- close enough for the formation of a merging BBH, but too wide for significant tidal spin-up of the He star.
    Through (8a,b) the second CC event, the system forms (9a,b) a BBH, which (10a,b) merges within the Hubble time. Due to the unusual stellar structure of the mass donor, a high-spin second-born BH might emerge (9b; see text). The values in grey parentheses show the component masses, in $M_\odot$, for two evolutionary examples (see text) selected from \citet{Xu2025arXiv251220054X}. Here, $M_{\rm BH,1st}$ and $M_{\rm BH,2nd}$ denote the masses of the first- and second-born BHs, and the ranges of $M_{\rm BH,2nd}/M_{\rm BH,1st}$ are intended only as approximate references. We refer to \citet{Xu2025arXiv251220054X} and \citet{Briel2026arXiv260203629B} for detailed binary evolution models on this channel. }
    \label{fig:close}
\end{figure*}

Figure\,\ref{fig:close} illustrates the formation paths of merging BBHs in initially close binaries through the SMT. 
During main-sequence evolution, the radius of a massive star can expand by a factor of a few, depending on stellar mass. Hence, in close binaries, the primary star can fill its Roche lobe during the MS phase (i.e., Case\,A mass transfer; 2 in Fig.\,\ref{fig:close}), which occurs with orbital periods below about 10\,d, depending on the primary mass \citep[Appendix\,A in][]{Xu2025arXiv250323876X}. Whereas Case\,B mass transfer strips most of the H-rich envelope of the post-MS donor in one episode (on thermal timescale), Case\,A systems have a more complex mass transfer history, which is usually treated in a simplified way in rapid population synthesis codes \citep[][and discussions in \citet{Olejak2024A&A...689A.305O}]{Marchant2024ARA&A..62...21M}. According to detailed binary evolution models \citep{Wellstein2001,Sen2022,Xu2025arXiv251220054X}, mass transfer in a Case\,A system firstly proceeds on the thermal timescale (fast Case\,A), usually making the mass donor less massive than the companion, and then continues on the nuclear timescale of the mass donor (slow Case\,A). Mass stripping during the MS phase reduces the size of the convective core of the mass donor \citep{Schurmann2024A&A...690A.282S,Xu2025arXiv251220054X}. Meanwhile, mass accretion rejuvenates the MS accretor, significantly altering the structure from a normal single star \citep{Braun1995A&A...297..483B,Renzo2021ApJ...923..277R,Xu2025arXiv251220054X}.
Once the mass donor depletes its core hydrogen, the rapid expansion triggers another phase of fast mass transfer (Case\,AB), transferring helium-rich material to the mass gainer. Observational counterparts of Case\,A systems have been identified at many key evolutionary stages, which are Algol systems, close WR star binaries, close BH+O binaries, WR star X-ray binaries \citep[][and references therein]{Xu2025arXiv251220054X}, suggesting that considerable mass accretion has occurred during the first MT phase, although how conservative it is remains unclear \citep{Petrovic2005,Sen2022,Nuijten2025A&A...695A.117N,Sen2026ApJ..1000....2S}.

After the first MT phase, the orbit remains relatively tight. To form a merging BBH, the second MT phase (6a,b in Fig.\,\ref{fig:close}) must remain stable to avoid the merger between the BH and its companion star. Only modest orbital shrinkage is required for the resulting BBH to merge within the Hubble time, in contrast to formation pathways originating from wide binaries \citep{Marchant2021,Gallegos-Garcia2021,Briel2023MNRAS.520.5724B,Xu2025arXiv251220054X,Briel2026arXiv260203629B,Klencki2026A&A...706A.296K}. 
In Fig.\,\ref{fig:close}, we present two evolutionary examples based on the detailed binary-evolution models in \citet{Xu2025arXiv251220054X}, which continuously evolve from the zero-age main sequence until the formation of the BBH. In the presented examples, the first MT is always Case\,A type (step 2 in Fig.\,\ref{fig:close}), while the second MT can be either Case\,A or Case\,B (step 6a,b in Fig.\,\ref{fig:close}), referred to as Case\,A-Case\,A and Case\,A-Case\,B types in \citet{Xu2025arXiv251220054X}.
\begin{itemize}
    \item The first example (labelled `a' in Fig.\,\ref{fig:close}) starts with a $32\mso$ primary and a $28\mso$ secondary \citep[the Case\,A--Case\,B example in][]{Xu2025arXiv251220054X}. After the Case\,A MT, the primary star becomes a $13\mso$ He star, while the secondary star grows to about $42\mso$. Assuming direct collapse \citep{Xu2025arXiv251220054X}, the first-born BH is expected to be $13\mso$. Meanwhile, the companion star of the BH (i.e., the mass donor of the second MT)
    is more compact than a normal single star of the same mass due to the accretion of helium-rich material during the prior Case\,A MT phase.
    Consequently, when the second MT occurs, the mass donor 
    has already depleted its core hydrogen, triggering a Case\,B MT phase, even though the orbital period is still below 10\,d (step 6a in Fig.\,\ref{fig:close}).
    Then, the mass donor becomes a $19\mso$ He star. Tidal spin-up during the following BH+He stage is limited. In the end, the system produces a merging BBH with a mass ratio near 0.7 but a low effective spin.
    \item The second example (labelled `b' in Fig.\,\ref{fig:close}) starts with a slightly lower initial mass ratio: a $32\mso$ primary and a $25\mso$ secondary \citep[the Case\,A--Case\,A example in][]{Xu2025arXiv251220054X}. The secondary star grows up to $39\mso$ after the first MT, and it fills its Roche lobe during the MS phase at the onset of the second MT. This second Case\,A MT exposes the helium-rich material above the convective core (step 6b in Fig.\,\ref{fig:close}). While the donor star is still on the MS after the second MT, its surface helium mass fraction exceeds 0.8, making it behave like a chemically homogeneous star (cf., Sect.\,\ref{sec:che}) and potentially produce a high-spin second-born BH. In this example, the resulting BBH has a near-unity mass ratio and a high-spin second-born BH \citep[see also][]{Ma2026arXiv260811311M}. 
\end{itemize}
Therefore, the SMT channel in close binaries is unlikely to produce BBH mergers with mass ratios below 0.5.
While the Case\,A--Case\,A models might produce high-spin second-born BHs, the mass ratios of the resulting BBHs are likely to exceed 0.8 \citep{Xu2025arXiv251220054X,Briel2026arXiv260203629B}. 
One caveat is BH kicks, which may produce lower mass ratios \citep{Briel2026arXiv260203629B}.

In addition, the above examples assume that the accretion onto a BH is limited by the Eddington rate, resulting in a near-zero accretion efficiency during the second MT (step 6a,b in Fig.\,\ref{fig:close}).
However, the first-born BH can be
significantly spun up if super-Eddington accretion occurs \citep[e.g.,][]{Zevin2022ApJ...933...86Z,Shao2022ApJ...930...26S,Briel2023MNRAS.520.5724B,Xing2025A&A...693A..27X,Briel2026arXiv260727962B}. 
In particular, producing a spin parameter above 0.5 for the first-born BH requires the second MT to be nearly conservative \citep{Xing2025A&A...693A..27X,Briel2026arXiv260727962B}.

\subsection{Close binaries: chemically homogeneous evolution\label{sec:che}}

\begin{figure*}[t]
    \centering
    \includegraphics[trim=0.7cm 9.4cm 5.5cm 0cm,clip,width=0.9\linewidth]{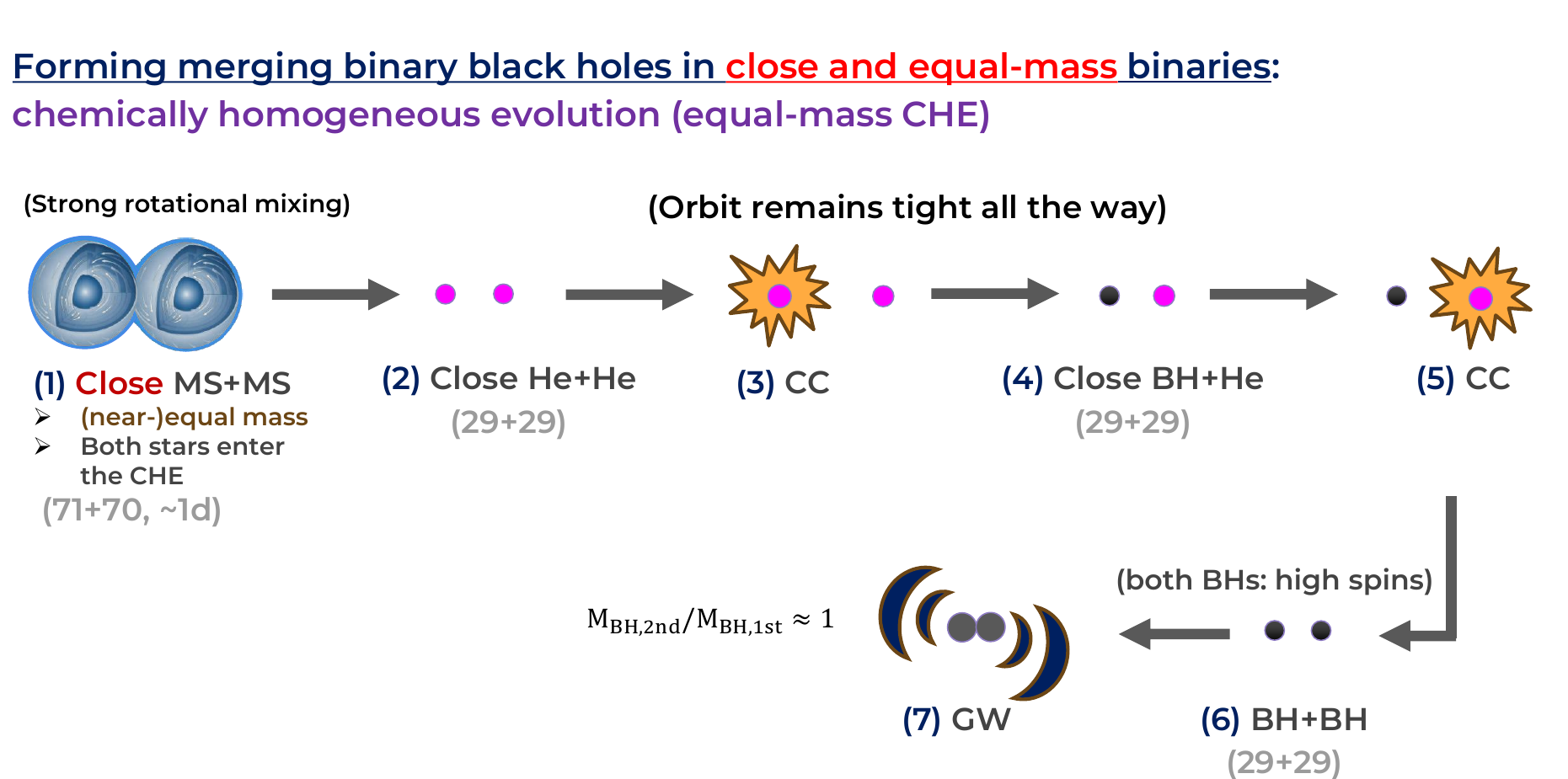}
    \caption{Formation path of merging binary black holes in initially close binaries if both stars evolve chemically homogeneously. The system begins with (1) a close binary containing two main-sequence (MS) stars of (near-)equal mass. Due to tidal spin-up and deformation, strong mixing occurs in both stars causing them to enter the chemically homogeneous evolution (CHE). The system evolves into (2) a double helium (He)-star binary. Then, it evolves through (3) the first core-collapse (CC) event, (4) a BH+He binary stage, and (5) the second CC event to produce (6) a BBH containing two high-spin BHs with (near-)equal masses. Finally, the BBH (7) merges within the Hubble time. 
    The values in grey parentheses show the component masses, in $M_\odot$, for an evolutionary example (see text), except that `1\,d' indicates the initial orbital period. Here, $M_{\rm BH,1st}$ and $M_{\rm BH,2nd}$ denote the masses of the first- and second-born BHs, and the ranges of $M_{\rm BH,2nd}/M_{\rm BH,1st}$ are intended only as approximate references. We refer to \citet{Marchant2016} for detailed binary evolution models on this channel and \citet{Mandel2016MNRAS.458.2634M} for population synthesis predictions.}
    \label{fig:CHE}
\end{figure*}

\begin{figure*}[t]
    \centering
    \includegraphics[trim=0.7cm 9.5cm 4.2cm 0cm,clip,width=0.95\linewidth]{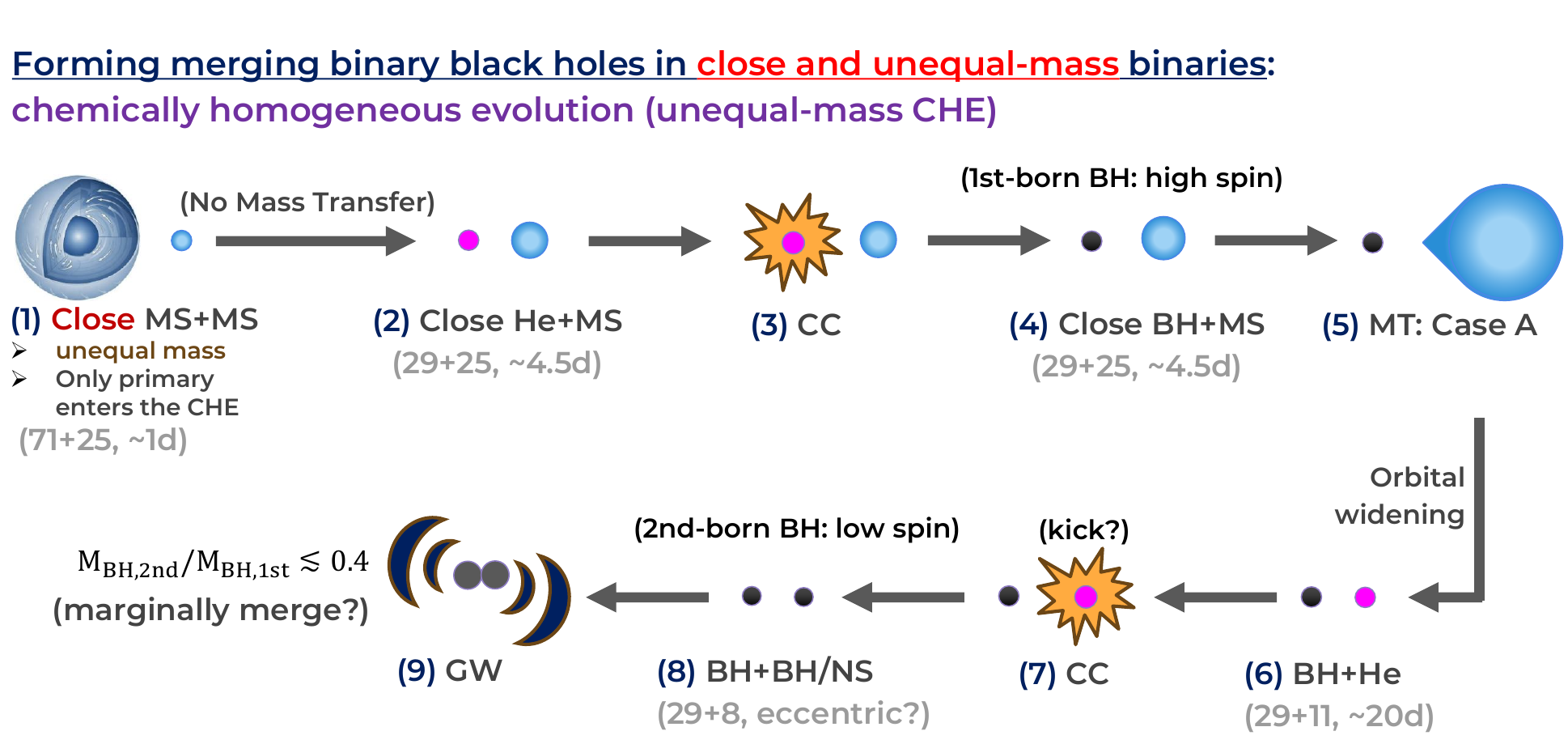}
	\caption{Formation path of merging binary black holes in initially close binaries if only the primary star evolves chemically homogeneously. The system begins with (1) a close binary containing two main-sequence (MS) stars of unequal masses. The efficiency of rotational mixing varies with stellar mass, allowing the primary star to enter the chemically homogeneous evolution (CHE) while the secondary star evolves normally. Since the primary star does not expand, the system evolves into (3) a close helium (He)-star+MS binary without mass transfer. Then, the He star can form (4) a high-spin first-born BH. Since the orbital period is still below 10\,d, the expansion of the initial secondary star triggers (5) Case\,A MT onto the first-born BH, during which orbital widening occurs, as the BH is likely more massive than the mass donor. After that, the system evolves into
	(6) a relatively wide BH+He binary. Through (7) the second CC event, the He star produces a low-spin second-born BH or a neutron star (NS). If the second CC is associated with a suitably oriented natal kick, the resulting BBH may merge within the Hubble time. 
    The values in grey parentheses show the component masses, in $M_\odot$, and orbital period for an evolutionary example (see text). Here, $M_{\rm BH,1st}$ and $M_{\rm BH,2nd}$ denote the masses of the first- and second-born BHs, and this formation path gives $M_{\rm BH,2nd}/M_{\rm BH,1st}\lesssim0.4$. 
	We refer to \citet{Marchant2017} for detailed binary evolution models on this channel and \citet{Xu2025arXiv250323876X} for population synthesis predictions for the BH+MS stage expected by this channel.}
    \label{fig:CHE-2}
\end{figure*}

Stellar rotation can induce various instabilities, causing mixing of stellar material \citep[e.g.,][]{Brott2011A&A...530A.116B,Jin2024A&A...690A.135J,Sun2025NatCo..16.9729S}. If hydrogen in the envelope is mixed into the convective core faster than it is consumed by nuclear burning, the star does not expand and directly evolves into a helium star.
This is known as the chemically homogeneous evolution (CHE), which
can produce high-spin BHs with spin parameters up to 1 \citep{Yoon2005,Woosley2006ApJ...637..914W,Marchant2016}, although this outcome is subject to the uncertainties in WR star winds \citep{Marchant2024A&A...691A.339M}. 
In close massive binary systems, CHE can be induced by tidal spin-up and deformation, potentially forming high-spin merging BBHs \citep{Marchant2016,Mandel2016MNRAS.458.2634M,Hastings2020A&A...641A..86H}. This pathway favours very massive stars, initial orbital periods of about 1\,d, and metallicities below about 1/10th solar value, such that rotational mixing is sufficiently strong and is not suppressed by wind braking \citep[e.g.,][]{Marchant2016}. In such tight orbits, most of the massive binaries will likely  enter an overcontact phase. Surviving this phase requires a relatively high initial mass ratio (i.e., $q$ near 1) to avoid overflow through the outer Lagrangian point. During this phase, the mass ratio rapidly approaches unity. The overcontact system, VFTS\,352, is a likely observational counterpart of this evolutionary path \citep{Almeida2015ApJ...812..102A}. 
Figure\,\ref{fig:CHE} presents a schematic of the CHE channel. We also provide an example, with masses indicated by numbers in grey parentheses. According to the detailed models computed with the metallicity of the Small Magellanic Cloud \citep{Xu2025arXiv250323876X}, the CHE occurs in close binaries with the initial primary star from about $70\mso$, forming a He core of $29\mso$. Therefore, we consider a close binary of  a $71\mso$ primary and a $70\mso$ secondary with an orbital period of about 1\,d, which produces an equal-mass double-He-star binary ($29\mso$+$29\mso$). Accordingly the system ends up as an equal-mass merging BBH, in which both BHs have high spins. Thus,
this formation path is unlikely to produce the unequal-mass, high-spin population inferred in GWTC-5.

Given that the efficiency of rotational mixing varies with stellar mass, it is possible that only the primary star evolves chemically homogeneously, while the secondary star evolves normally \citep{Marchant2017}. Figure\,\ref{fig:CHE-2} presents a schematic of this evolutionary sequence, together with an example. The initial parameter of the example is selected from Appendix\,G.2 of \citet{Xu2025arXiv250323876X}. The evolution starts with a $71\mso$ primary, a $25\mso$ secondary, and a period around 1\,d. Since the primary star does not expand in this case, it forms a high-spin BH of $29\mso$ without filling its Roche lobe. The expansion of the initial secondary star triggers a Case\,A MT, causing considerable orbital widening from about 4.5\,d to about 20\,d (assuming isotropic mass loss and Eddington-limited accretion). The initial secondary produces an $11\mso$ He star, likely ending up with a second-born BH of $8\mso$. If the second-born BH receives a suitably oriented natal kick, the resulting BBH could merge within the Hubble time, resulting in a highly unequal-mass BBH merger with a high-spin primary BH.

\section{Spin-orbit misalignment\label{sec:misalignment}}

Our survey in Sect.\,\ref{sec:channels} shows that there are two reasonable pathways to form unequal-mass BBHs with a high primary spin, as illustrated in Figs. \ref{fig:wide} and \ref{fig:CHE-2}.
Among the high-spin merging BBHs in GWTC-5, GW241110 stands out because of the strong evidence for a large spin-orbit misalignment \citep{Abac2025ApJ...993L..21A}, while the spin-orbit misalignments remain poorly constrained for other events. Due to the effects of tides and mass transfer, the BH progenitors in isolated binaries are expected to have spins aligned with the orbital angular momentum \citep{Mandel2022PhR...955....1M}. Hence, producing a large misalignment in isolated binaries generally requires a large natal kick and favourably oriented natal kick, with a magnitude at least comparable to the binary's relative orbital velocity when the second-born BH forms.

Consider a post-CEE binary containing a $20\mso$ He star and an $8\mso$ BH in a circular orbit with a period of 1\,d (i.e., 7a in Fig.\,\ref{fig:wide}). The relative orbital velocity is about $650\,\kms$, which is significantly larger than the current expectation for BH kicks produced in supernovae \citep[e.g.,][]{Fragos2009ApJ...697.1057F,Wong2012ApJ...747..111W,Mandel2016MNRAS.456..578M,Vigna-Gomez2024PhRvL.132s1403V}. 
However, in the unequal-mass CHE channel (Fig.\,\ref{fig:CHE-2}), the orbit widens considerably during the MT triggered by the initial secondary star. In the provided example, an $11\mso$ He star orbits around a $29\mso$ high-spin BH with a period of about 20\,d (step 6 in Fig.\,\ref{fig:CHE-2}). The corresponding relative orbital velocity is about $270\,\kms$, which may be achievable for BH kicks \citep[e.g.,][]{Fragos2009ApJ...697.1057F}. If such a kick is suitably oriented, the system may produce an unequal-mass merging BBH with a misaligned, high-spin primary BH.

Given that about 30\% of massive stars are expected in triple systems \citep{Moe2017ApJS..230...15M}, we expect some merging BBHs to have a tertiary companion. A tertiary companion can induce spin-orbit misalignment through nodal precession of the inner orbit ($\boldsymbol{L}_\mathrm{in}$ around $\boldsymbol{L}_\mathrm{out}$) as long as the spin of the primary BH ($\boldsymbol{S}$) does not adiabatically follow the orbital angular momentum ($\boldsymbol{L}_{\rm in}$). For this to happen, the period of nodal precess ($P_{\rm prece}$) must be shorter than the period of the de Sitter precession of $\boldsymbol{S}$ around $\boldsymbol{L}_\mathrm{in}$, and $P_{\rm precess}$ must also be shorter than the merger time of the inner BBH \citep{Liu2017ApJ...846L..11L}.  
Assuming modest eccentricities for the inner and outer orbits, $P_{\rm prece}$ is estimated as
\begin{equation}
\begin{split}
    P_{\rm prece} &\approx P_{\rm in} \left[\frac{3}{8}\frac{M_3}{M_{\rm 1}+M_2}\left(\frac{a_{\rm in}}{a_{\rm out}}\right)^3\right]^{-1}\\
    &\approx\frac{8}{3} \frac{M_1+M_2+M_3}{M_3}\frac{P_{\rm out}^2}{P_{\rm in}},
        \label{eq:p-prece}
\end{split}
\end{equation}
where $P_{\rm in}$ ($P_{\rm out}$) is the orbital period of the inner binary (tertiary companion), $a_{\rm in}$ ($a_{\rm out}$) is the semi-major axis of the inner binary (tertiary companion), $M_3$ is the mass of the tertiary companion, and
$M_1$ ($M_2$) is the mass of the primary (secondary) BH of the inner binary.
The period of the de Sitter precession ($P_{\rm SL}$) of $\boldsymbol{S}$ around $\boldsymbol{L}_{\rm in}$ is given by
\begin{equation}
\begin{split}
    P_{\rm SL} &\approx P_{\rm in} \left[\frac{3}{2}\frac{G (M_2+\mu/3)}{c^2a_{\rm in}}\right]^{-1}\\
    &\approx 4\times 10^6\,\mathrm{day}\, \left(\frac{P_{\rm in}}{\rm day}\right)^{5/3} \frac{(m_1+m_2)^{4/3}}{m_2(4m_1+3m_2)},
\end{split}
\end{equation}
where $\mu$ is the reduced mass of the inner binary, $m_1 = M_1/\mso$, and $m_2=M_2/\mso$.  

A tertiary companion can also induce Zeipel-Lidov-Kozai (ZLK) oscillations \citep{Zeipel1910AN....183..345V,Kozai1962AJ.....67..591K,Lidov1962P&SS....9..719L}, which excite the eccentricity of the inner binary and potentially shorten the merger time \citep{Antonini2017ApJ...841...77A,Liu2018ApJ...863...68L,Vigna-Gomez2025A&A...699A.272V,Stegmann2025ApJ...991L..54S}.  Such oscillations occurs on the same timescale as $P_\mathrm{prece}$, and can operate efficiently if the inner binary is not too compact, as relativistic precession can suppress the eccentricity excitation by ZLK oscillations \citep{Liu2018ApJ...863...68L}. This condition can be described by the $\varepsilon_{\rm GR}$ parameter defined in \citet{Liu2018ApJ...863...68L}, which is 
\begin{equation}
\begin{split}
    \varepsilon_{\rm GR}& = \frac{3G(M_1+M_2)^2a_{\rm out}^3}{c^2a_{\rm in}^4M_3}\\
    & \simeq 1.5\times 10^{-6} (m_1+m_2)^{2/3}\frac{m_1+m_2+m_3}{m_3}\\
    &\times\frac{(P_{\rm out}/\text{day})^2}{(P_{\rm in}/\text{day})^{8/3}}, 
    \label{eq:vare-gr}
\end{split}
\end{equation}
where $m_3=M_3/\mso$, and the orbit of the tertiary is assumed to be circular. ZLK oscillations require $\varepsilon_{\rm GR}\lesssim 1$. For comparable-mass inner binaries, $\varepsilon_\mathrm{GR} \sim P_{\rm prece}/P_{\rm SL}$, so the condition $\varepsilon_\mathrm{GR} \lesssim 1$ also implies that significant spin-orbit misalignment may be generated.

Considering a circular inner binary of a $20\mso$ primary BH and an $8\mso$ secondary BH with a period of one day (see Fig.\,\ref{fig:wide}, step 9a,b), the merger time is about 0.9\,Gyr. According to Eq.\,\eqref{eq:p-prece}, we have $P_{\rm prece}< 0.9\,$Gyr if a $20\mso$ tertiary has its orbital period $P_{\rm out}\lesssim 2\times10^5\,$d (590\,yr). Meanwhile, the condition of $P_{\rm prece} < P_{\rm SL}$ requires $P_{\rm out}\lesssim 260\,$d. However, mass transfer in isolated massive binaries can occur with initial orbital periods (i.e., the period of step 1 in Fig.\,\ref{fig:wide}) up to a few $10^3$\,d \citep{Xu2025arXiv250323876X}, exceeding the estimated upper limit of $P_{\rm out}$. Therefore, a tertiary companion can hardly produce spin-orbit misalignment for the pathways in initially wide binaries (see Fig.\,\ref{fig:wide}). Otherwise, the tertiary would need to be initially wide enough to avoid affecting the evolution of the inner binary, but subsequently migrate inward to satisfy the $P_{\rm out}\lesssim260\,{\rm d}$ condition.

We now analyse whether a tertiary companion can generate spin-orbit misalignment for the unequal-mass CHE channel (Fig.\,\ref{fig:CHE-2}).
The inner binary is initially compact in order to trigger the CHE, so that ZLK oscillations are suppressed, avoiding a premature binary merger. In the later MT phase  (step 5 in Fig.\,\ref{fig:CHE-2}), the orbit widens, 
potentially allowing ZLK oscillation to operate. For a BBH containing BHs of $29\mso$ and $8\mso$ with a period of 20\,d (step 8 in Fig.\,\ref{fig:CHE-2}), ZLK oscillations can operate if a $20\mso$ tertiary has an orbit period below about $20\,$yr according to Eq.\,\eqref{eq:vare-gr}. Thus, a tertiary could potentially enhance the merger rate of the unequal-mass CHE channel and also produce a large spin-orbit misalignment.

\section{Conclusion\label{sec:conclusion}}

With the release of GWTC-5, a subpopulation of merging BBHs, featuring a high-spin primary BH and a preference for unequal BH masses, has emerged \citep{LVK2026arXiv260527225T}. 
In this paper, we have analysed the possible formation scenarios for such BBHs through isolated binary evolution. We have identified two possible formation pathways (Figs.\,\ref{fig:wide} and \ref{fig:CHE-2}). 

The formation paths of merging BBHs in initially wide binaries requires strong orbital shrinkage to bring the binary components close together, which can occur through either CEE or SMT (step 6a,b in Fig.\,\ref{fig:wide}). Prior to the final BBH stage, the binary evolves through a tight BH+He phase, during which tidal spin-up can lead to a high-spin second-born BH. Here, the second-born BH can be more massive than the first-born BH if the initial secondary star accretes a large amount of material during the first MT phase, or if a strong mass ejection occurs during the formation of the first-born BH, resulting in an unequal-mass BBH with a high-spin primary BH. 

In unequal-mass close binaries, the primary star can evolve chemically homogeneously, while the secondary star evolves normally (Fig.\,\ref{fig:CHE-2}). The primary star produces a high-spin first-born BH without transferring mass to the secondary star, and the second-born BH is likely less massive than the first-born one. However, mass transfer from the initial secondary star widens the orbit. Thus, the resulting BBH might merge within the Hubble time if the second-born BH forms with a considerable, suitably oriented natal kick.

GW241110 exhibits a large spin-orbit misalignment, which is usually interpreted as the result of hierarchical merger \citep{Abac2025ApJ...993L..21A}. In our isolated binary evolution scenarios, spin-orbit misalignment can be produced by natal kicks or by a tertiary companion (nodal precession or ZLK oscillations). If the spin of the primary BH is generated by the tidal spin-up during the BH+He phase (step 7a,b in Fig.\,\ref{fig:wide}), reproducing the misalignment would require a very large BH kick or a sufficiently close tertiary that might also affect the evolution of the progenitor binary of the inner BBH. If the high spin is the outcome of  CHE (the unequal-mass CHE; Fig.\,\ref{fig:CHE-2}),  the required BH kick may be
achievable, and the orbital widening in later evolution also opens a window for ZLK oscillations.  

Overall, our analysis suggests that hierarchical mergers are not uniquely required for the formation of unequal-mass BBHs with a high-spin primary BH, and alternative scenarios based on isolated binary evolution can also produce such BBHs. The evolution of isolated binaries and triples generally predict a wide range of the spin of the primary BH, while the product of a comparable-mass BBH merger naturally have spins near 0.7. Future detections of more events with spin around 0.7 would  discriminate between isolated binary evolution and hierarchical mergers.

\begin{acknowledgments}
X.-T.X. was supported by the Tsung-Dao Lee postdoctoral fellowship at the Tsung-Dao Lee Institute (TDLI) and the National Natural Science Foundation of China (grant No. 12603055). B.L. acknowledges support from the National Natural Science Foundation of China (grant No. 12673087 and 12433008) and National Key Research and Development Program of China (No. 2023YFB3002502)
\end{acknowledgments}

\bibliography{Xu_SMC}
\bibliographystyle{aasjournal}

\end{CJK*}

\end{document}